\documentclass[aps,prd,showpacs]{revtex4}
\usepackage{graphicx}
\usepackage{epsfig}
\usepackage{psfig}
\usepackage{revsymb}
\begin{document}
\renewcommand{\thesection}{\arabic{section}}
\renewcommand{\thesubsection}{\arabic{subsection}}

\title{RHIC Perfect Fluid sQGP- a Landau Chromo-Diamagnetic Liquid}
\author{Somenath Chakrabarty}
\affiliation{Department of Physics, Visva-Bharati, Santiniketan 731 235, 
West Bengal, India, E-mail:somenath@vbphysics.net.in}
\begin{abstract}
We speculate that the quark matter produced at RHIC, behaves like a
perfect fluid may be considered as Landau chromo-diamagnetic material in 
presence of
strong color magnetic field. We have shown that in such a system the
viscosity coefficient can be small enough and in the extreme case of
ultra-strong color magnetic field strength, it behaves like a super fluid. 
\end{abstract}
\maketitle
The possibility of deconfinement transition in ultra-relativistic
heavy-ion collisions was predicted long ago. A lot of theoretical work
have been done on the order of phase transition to quark matter and also
the possible signatures of QGP formation in ultra-relativistic heavy-ion
collisions. A number of models for space-time evolution of QGP
phase were also proposed \cite{R1}.
In the cosmological scenario, it is also expected that the universe was
filled with hot and dense soup of QGP at about micro-second after
big-bang \cite{R2}.

In the past high energy heavy-ion collision experiments, starting from SPS
energies (which is of course not too high), the theoretically predicted 
experimental signatures of QGP were either not observed or were found
anomalous in the sense that the same signatures one could get if it were
instead of QGP a hot and dense hadronic matter.
The main reason behind all these ambiguities is the
scale of incident energy of the heavy-ion beam. 

Only recently RHIC offers the experimentalists and also the theoreticians
who once predicted a lot of experimental signatures of QGP to study such a
hot and dense system in the extreme situation, which was also expected to be
present in the very early universe. At RHIC, when two gold nuclei moving
with almost the speed of light, collide, the amount of energy that dump in
the microscopic fireball is about $2.0\times 10^4$GeV. The pressure
generated at the time of impact of two gold nuclei is truly immense, it is
about $10^{30}$ atmospheric pressure and the temperature inside the
fireball is about trillions of degrees. Under such extreme condition, it is
expected that all the (constituent) quarks of the colliding nuclei should
come out of their parent hadrons and form a gas of hot and dense quark
matter. Not only that, because of such high temperature and densities, an
enormous number of quark anti-quark pairs are also produced from vacuum
and all of them move in a self-consistently generated color field. Bcause
of pair annihilation processes within the hot QGP system, there will be a
lot of thermal gluons. This whole thing is expected to behave like a
plasma of quarks (anti-quarks) and gluons, called quark-gluon plasma
(QGP). This physical scenario was predicted long ago by the theoretical
high-energy heavy-ion physicists or high-energy nuclear physicists.

Surprisingly, the physical picture emerging from the detectors BRAHMS,
PHENIX, PHOBOS and STAR of four different experimental groups is consistent.
The recorded data of sub-atomic debris coming out of the violent collisions
at the four colliding points of RHIC, observed by the four detectors as
mentioned above, show a complete deconfinement of quarks (constituents and
sea quarks) and gluons. Most surprisingly, this hot and dense quark matter, 
instead of a gas,
behaves like a liquid in thermal equilibrium. It exhibits a collective flow of  
almost a perfect
liquid, i.e., a fluid of extremely low shear viscosity \cite{R3}. The elliptic 
flow of
QGP is a direct evidence of its liquid like behaviour. Further, the elliptic
pattern of flow indicates the presence of substantial pressure gradient in
QGP and that the quarks and gluons which subsequently produce hadrons having
anisotropic spatial distribution, show the collective nature.
People are calling this quark matter system by strongly correlated QGP or in
short, sQGP.

This situation conflicts with the theoretical prediction that the quark
matter if produced in ultra-relativistic heavy-ion collisions will be almost
ideal, weakly interacting gas. Since the hadrons from RHIC sQGP are not
emitted uniformly in all possible directions, it is not acceptable that the
hot quark matter is in a gas phase. On the other hand, it is seen that
the produced quark matter behaves like a most perfect liquid ever observed.

There are a number of theoretical attempts and also some suggestions to
explain this surprising physical scenario produced at RHIC. One of the
approach is the lattice QCD calculation using high speed sophisticated
computers. However, the most up to date version of lattice QCD calculation is
unable to handle the rapidly changing scenario of RHIC sQGP and reproduce
the jet quenching and low viscosity phenomenon \cite{R4}. It is also suggested 
to get
some help from string theorists, people working on super-symmetry,
super-gravity, quantum black holes, etc. In a recent work an anomalous
viscosity \cite{R5}, which is low enough compared to kinetic viscosity is introduced 
for expanding RHIC sQGP. It is shown using the turbulent motion of QGP in
presence of self-consistently generated color magnetic fields that there can 
exist an
anomalous viscosity of very low magnitude compared to the viscosity arising
from particle collisions.

In the present article we shall try to explain the low viscosity nature of
RHIC sQGP from a completely different line of thought. We assume the motion
of thermal gluons and quarks (anti-quarks) in a self consistently generated
color magnetic field. Now it is well known from QED that in a charge particle 
system in presence of an external magnetic field, if 
the strength of the field exceeds some quantum critical value, which is a 
function of mass
and magnitude of charge carried by the particle concerned, the Landau levels
of the charge particles will be populated \cite{R6}. The phenomenon is called 
Landau diamagnetism. It is the quantum mechanical effect of strong magnetic field.
The phenomenon is also known as Landau quantization (the transverse component of
particle momentum gets quantized) in condensed matter physics. We expect that the 
same
phenomenon can be possible in presence of self-consistently generated color
magnetic field inside RHIC sQGP. The quarks (anti-quarks) and thermal gluons, 
carrying color charges 
will interact with the strong color magnetic field quantum mechanically and 
if the field strength exceeds the corresponding quantum critical limit, 
the Landau levels of these particles will be populated
and as a consequence the RHIC sQGP will behave like a Landau chromo-diamagnetic 
system. To develop a formalism to obtain the shear viscosity coefficient of
such a system, the first step we have followed is to choose the gauge
$A_a^\mu\equiv (0,0,xB_a,0)$, where $A_a^\mu$ is the color electric field
vector and $B_a$ is the color magnetic field along $z$-axis, $a=1,2,..8$,
the color index. Then the single particle energy for quarks (anti-quarks) 
is given by
\begin{equation}
E_{n,\alpha}(k_z)=\left [ k_z^2+m^2+2eB(n+\frac{1}{2}+\alpha)\right ]^{1/2}
\end{equation}
where $k_z$ is the $z$-component of quark momentum, varies continuously from
$-\infty$ to $+\infty$, $m$ is the quark mass, $n=0,1,2,..$ is called 
Landau quantum
number, $\alpha=\pm 1/2$ the spin projection and $eB=\sum gq_aB_a$, with $q_a$ is 
the color charge related to the generators of irreducible $SU(3)$
representation and $g$ is the corresponding strength and the sum is over
the color index. The quantity $[(2n+1+2\alpha)eB]^{1/2} =p_\perp$ is the 
orthogonal component of quark (anti-quark)
momentum. Since the transverse part of momentum $p_\perp$ is changing in a 
discrete manner, the phenomenon is  called the Landau
quantization. We can similarly write down the energy eigen value for gluons,
given by
\begin{equation}
E_{n,\alpha}(k_z)=\left [ k_z^2+2eB(n+\frac{1}{2}+\alpha)\right ]^{1/2}
\end{equation}
where, in this case $\alpha=\pm 1$, the corresponding spin projections for two
physical degrees of freedom of gluons. In the present scenario the
geometrical structure of momentum space for the colored particles will also 
get modified. It will become cylindrical instead of usual spherical
structure, with the axis along the direction of color magnetic field.

Now it is already known from QED calculations that the upper limit of Landau
quantum number ($\nu_{\rm{max}}$) for quarks decreases with the increase of 
magnetic field strength. Therefore only low lying Landau levels will be
populated for quarks (of course there will be no change for the gluonic
case, which are vector bosons carrying color charges) if the strength of 
color magnetic field is strong enough. In the
extreme situation, when $\nu_{\rm{max}}$ becomes zero, all the quarks
(anti-quarks) will occupy only the zeroth Landau level. In this case the
effect of color magnetic field on (color) charged particles will be most
prominent.  

To obtain an expression for shear viscosity coefficient for expanding RHIC
sQGP, we study the transport theory of this Landau chromo-diamagnetic matter
following the relativistic version of Boltzmann kinetic equation. The relevant 
part of which is given by \cite{R7}
\begin{equation}
p^\mu\partial_\mu f(x,p)=C[f]
\end{equation}
where $f(x,p)$ is the distribution function and $C[f]$ is the collision
terms, which contains the rates of all possible elementary processes. With
the relaxation time approximation, we have
\begin{equation}
C[f]=-\frac{p^0}{\tau}[f(x,p)-f_0(p)]
\end{equation}
where $\tau$ is the relaxation time and $f_0(p)$ is the equilibrium
distribution function (Fermi distribution for quarks and anti-quarks and Bose
distribution for gluons). We further make linear approximation for the
non-equilibrium distribution function, which gives
\begin{equation}
f(x,p)=f_0(p)[1+\chi(x,p)]
\end{equation}
Then following de Groot \cite{R7}, we have the expression for shear viscosity
coefficient for quarks (anti-quarks)
\begin{eqnarray}
\eta_{q\bar q}&=&\frac{g_q}{T}\sum_{\nu=0}^{\infty} (2-\delta_{\nu
0})\frac{eB}{4\pi^2} \int_{-\infty}^{+\infty} \frac{\tau_q}{E_\nu}dp_z
\nonumber \\ && (p_rp_z)^2
\left [ f_0(p)(1-f_0(p))
+\bar f_0(p)(1-\bar f_0(p))\right ]
\end{eqnarray}
where $q_g=(2N_f+1)(2N_c+1)$ is the degeneracy factor, $T$ is the
temperature of the system, $p_r=p_\perp=(2\nu eB)^{1/2}$ and
$2\nu=(2n+1+2s_z)$, with $\nu=0,1,2,..$ etc. 

Similarly for gluons, we have
\begin{equation}
\eta_g=\frac{g_g}{T}\sum_{\nu=0}^{\infty} (2-\delta_{\nu
0})\frac{eB}{4\pi^2} \int_{-\infty}^{+\infty} \frac{\tau_g}{E_\nu}dp_z
 (p_rp_z)^2
f_0(p)(1+f_0(p))
\end{equation}
where the degeneracy factor $g_g=(2N_c+1)$. 
It is easy to show that that the
zeroth Landau levels for quarks, anti-quarks and gluons are singly degenerate, 
whereas all other states are
doubly degenerate.
To obtain these expressions, we
have assumed that the flow is along $z$-direction. However, since the
expressions are symmetric with respect to $p_r$ and $p_z$, there will be no
qualitative or quantitative change if we assume flow along some direction on
the $x-y$ plane. Now following the arguments that the relaxation rates are
additive, we have
\begin{equation}
\frac{1}{\eta_s}=\frac{1}{\eta_{q\bar q}}+\frac{1}{\eta_g}
\end{equation}
where $\eta_s$ is the effective viscosity coefficient of the system. It is
quite obvious from the expression for shear viscosity coefficient for quarks
(anti-quarks) that in the extreme case when only the zeroth Landau level
is occupied, the system behaves like a super fluid. In fig.(1), just for
illustration we have
shown the variation of effective viscosity coefficient with the strength of
color magnetic field, normalized with respect to QED quantum critical value
for electron ($4.4\times 10^{13}$G). We have chosen
two different temperatures and densities of the matter. To obtain the relaxation
time, we have considered the following elementary processes: 
$qq\rightarrow qq$,
$q\bar q\rightarrow q\bar q$,
$\bar q\bar q\rightarrow \bar q\bar q$,
$qg\rightarrow qg$,
$\bar qg\rightarrow \bar qg$, and
$gg\rightarrow gg$.
We have obtained the relaxation times $\tau_q$ and $\tau_g$ for the
reactions involving quarks (anti-quarks) and gluons respectively by evaluating 
the rates of the processes. Then the relaxation times are obtain from the
well know formula
\begin{equation}
\frac{1}{\tau_m}=\sum_i\frac{1}{\tau_i}
\end{equation}
where $m=q$ or $g$ and the sum is over all possible processes.
As there are a lot of uncertainties in the parameters used, e.g., we have
taken the strong coupling constant $\alpha_s=0.8$, the densities and
the temperatures chosen are again arbitrary, the quantitative nature of the curves 
therefore should not be taken too seriously. However, we expect that the
qualitative nature of the curves will not be changed if one uses more exact
input parameters.

The figure shows that at low and moderate values of color magnetic field, the
shear viscosity coefficient is almost insensitive with the variation of
magnetic field strength. However, beyond some upper limit, when the quarks
(anti-quarks) occupy only a few low lying Landau levels, the shear viscosity
decreases and in the extreme case, when only the zeroth Landau levels are
occupied, it exactly vanishes and as a consequence the RHIC sQGP will behave
like a super-fluid. We believe that the strength of color magnetic magnetic
field will be several orders of magnitude larger than the QED quantum critical
value for electron and as an outcome, the system should behave like a
perfect liquid or super fluid in the extreme case. Then following 
Feynman we can say that QGP in the extreme case will behave like dry water. 
We further
expect that although the system is a perfect fluid, it can not be a good
color conductor. Just like the shear viscosity, the color conductivity will
either vanish or will be very small 
in presence of strong color magnetic field. As a consequence the
physical nature of the the system will be something like a perfect fluid
of bad color conductor or in the extreme case a color neutral super fluid,
i.e., the system can not behave like a color super conductor (super fluid of
particle with color charges). This will be of course the most surprising
phenomenon ever observed. Perhaps such strange physical scenario 
will be another direct proof
of strongly coupled collective behaviour of RHIC sQGP.
We further believe that in LHC also the same qualitative nature will be prevailed.

Finally we expect that although RHIC sQGP is a deconfined thermalized strongly 
correlated quark matter system having collective behaviour, the chiral symmetry 
will not be restored in the system. The presence of strong
color magnetic field will generate mass dynamically just like the QED case.

\begin{figure}
\psfig{figure=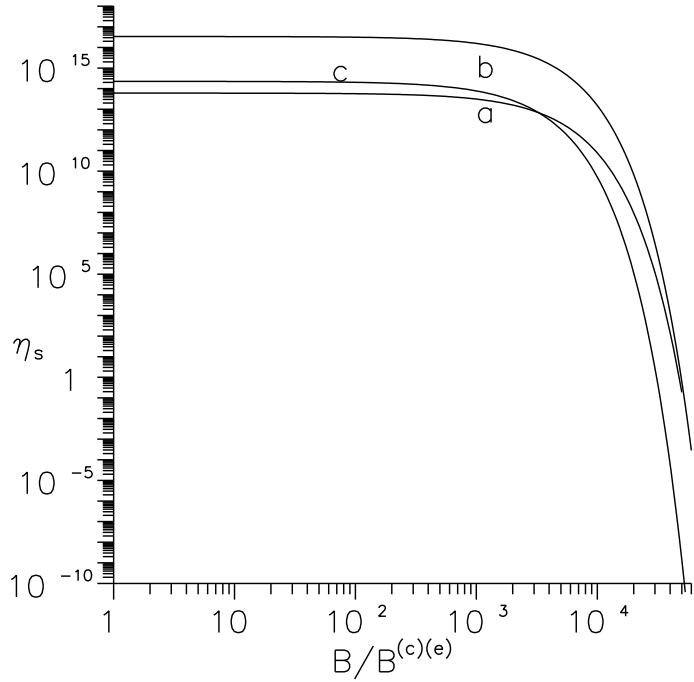,height=0.5\linewidth}
\caption{
Variation of shear viscosity coefficient with color magnetic field $B$
normalized with respect to the quantum critical value for electron.
Curve (a): $T=150$MeV and $n_B=5n_0$, Curve (b): $T=150$MeV and $n_B=8n_0$,
and Curve (c): $T=200$MeV and $n_B=5n_0$,
}
\end{figure}
\end{document}